\documentclass[a4paper,
               biblatex,     
               keeplastbox,   
               ]{jacow}
\usepackage{pdfpages,multirow,ragged2e} %
\usepackage{siunitx}

\makeatletter%
	\ifboolexpr{bool{xetex}}
	 {\renewcommand{\Gin@extensions}{.pdf,%
	                    .png,.jpg,.bmp,.pict,.tif,.psd,.mac,.sga,.tga,.gif,%
	                    .eps,.ps,%
	                    }}{}
\makeatother

\ifboolexpr{bool{xetex} or bool{luatex}} 
 {}                                      
 {\usepackage[utf8]{inputenc}}           

\usepackage[USenglish]{babel}

\ifboolexpr{bool{jacowbiblatex}}%
 {%
  \addbibresource{MOP010ref.bib}
 }{}
\begin{document}

\title{Emittance tuning of the FCC-ee High Energy Booster ring\thanks{This work was supported by the European Union’s Horizon 2020 Research and Innovation programme under grant no. 951754 (FCCIS).}}

\author{Q. Bruant\thanks{quentin.bruant@cea.fr}, B. Dalena\thanks{barbara.dalena@cea.fr}, 
		A. Chance, V. Gautard, CEA, Saint-Aubin, France \\
            F. Bugiotti, LISN, CentraleSupélec, CNRS, Paris-Saclay University, Saclay France\\
            A.Ghribi, GANIL, Caen, France\\
            R.Tomas Garcia, CERN, Geneva, Switzerland}
	
\maketitle

\begin{abstract}
    Previous studies for the electron-positron version of the Future Circular Collider (FCC-ee) have highlighted the need to define tolerances on magnet imperfections and develop correction strategies. This is crucial for ensuring the performance of one of the main elements in the acceleration chain: the High Energy Booster (HEB) ring.
    The efficiency and overall performance of these correction strategies, as well as the magnet field quality and misalignment tolerances, directly influence the specifications of correction magnets. This, in turn, affects key parameters such as beta functions, dispersion, transverse coupling, and emittance.
    Horizontal and vertical orbit corrections utilize horizontal and vertical kickers, respectively. Skew quadrupoles address vertical dispersion, introduced by normal dipole roll, and transverse coupling. Normal quadrupoles corrects the horizontal and vertical phase advances. This study simulates the distribution of these four corrector types to minimize equilibrium emittance at the extraction energy of 45.6 GeV. The computed strengths of these correctors and the associated misalignments are presented.
\end{abstract}

\section{INTRODUCTION}

The electron-positron version of the Future Circular Collider (FCC-ee) is one of the main proposals in development as CERN's next high-energy accelerator. The project envisions a 91-kilometer-long collider ring, accompanied by a second ring—the High Energy Booster (HEB)—housed within the same tunnel \cite{Benedikt:2928793}. To enable the practical manufacturing of all necessary components, it is essential to analyze imperfections in the magnetic lattice. This study focuses on defining tolerance levels and correction strategies in order to guarantee the target performances of the HEB ring. The procedure used is based on well known techniques and algorithms \cite{PhysRevAccelBeams.20.054801, BRANDT1990305, SAFRANEK199727}, but tailor-made for the distinctive features of the HEB.\\
In the following sections, the errors considered for this study are briefly discussed. Then, the choices made to implement the corrections necessary to limit the impact of these errors are presented. Finally, the results obtained following these choices are displayed.

\section{ERRORS}

Imperfections must be accounted for to ensure that the machine performs as expected and reaches its design performance. The field error and the positioning error were accounted for the HEB magnetic system including dipoles, quadrupoles, and sextupoles. The misalignment for the Beam Position Monitors (BPMs) is also accounted. These errors takes into account the misalignement of girders supporting these elements. Additionally, the relative field errors of the arcs dipoles, the arcs quadrupoles and the arcs sextupoles are considered. These errors are randomly assigned to the components using a Gaussian probability density function, truncated at three standard deviations (RMS). The RMS values for each type of error are provided in Table~\ref{Errors}.
\begin{table}[!hbt]
    \setlength\tabcolsep{3.5pt}
    \centering
    \caption{RMS value of the error distribution.}
    \label{Errors}
    \begin{tabular}{ccc}
        \toprule
        \textbf{Error Type} & \textbf{Unit} & $\sigma$\\
        \midrule
        Relative dipole field error & [1] & \num{e-3}\\
        Relative quadrupole field error & [1] & \num{2e-4} \\
        Relative sextupole field error & [1] & \num{2e-4}\\
        Main dipoles' roll error & [\unit{\micro\radian}] & 300 \\
        Main quadrupoles' roll error & [\unit{\micro\radian}] & 300\\
        Main girders' offset ($\sigma_{girder}$)& [\unit{\micro\metre}] &  200\\
        Main quadrupoles' offset & [\unit{\micro\metre}] & $\sigma_\text{girder} + 50$\\
        Main sextupoles' offset & [\unit{\micro\metre}] & $\sigma_\text{girder} +50$\\
        BPMs' offset & [\unit{\micro\metre}] & $\sigma_\text{girder} +50$ \\
        \bottomrule
    \end{tabular}
\end{table}
These RMS values are determined based on the ones published in \cite{errors} where the tolerances on girder-to-girder misalignment are a little bit relaxed in addition to the relative dipolar field error.

\section{CORRECTION STRATEGY}

\begin{figure}[!htb]
   \centering
   \includegraphics[width=.85\columnwidth]{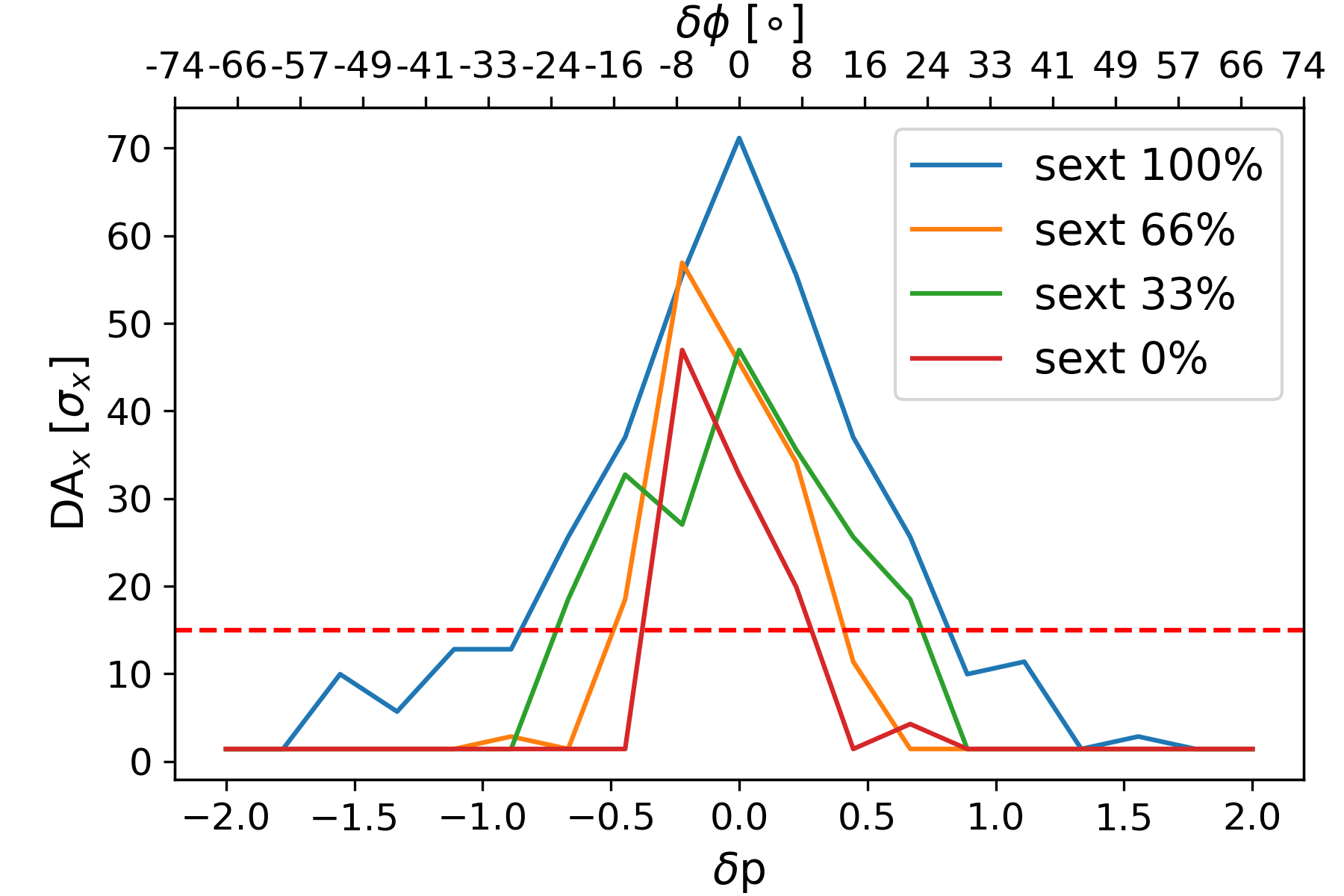}
   \caption{Momentum aperture as a function of energy spread for different sextupole strengths.}
   \label{fig:momap}
\end{figure}

The first step in the correction strategy is to establish a closed orbit in the simulated ring while accounting for previously defined errors. This approach is inspired by techniques used in existing accelerators.
The orbit correctors are assigned as follows: when a quadrupole focuses in the horizontal (or vertical) plane, the adjacent BPM measures in the same plane, and the next corrector, placed after the BPM, applies the correction in that same plane. The orbit correction process follows a method similar to that used during the LHC commissioning \cite{PhysRevSTAB.12.081002}, beginning with a segment-by-segment (SbS) correction—arc-by-arc in this case. After completing the correction on the first segment, each of the following segments is then added one by one to finally obtain the full lattice correction. This step enhances the stability of the correction process by effectively addressing a wider range of error configurations.
At this stage, an initial correction round is performed with the sextupoles turned off. The orbit obtained after SbS optimization undergoes two singular value decomposition (SVD) optimizations over the full lattice using the \textsc{MAD-X CORRECT} command \cite{MADX_doc} to minimize residual excursions. Typically, two SVD iterations are sufficient to determine the closed orbit. Further SVD optimizations are then applied to refine the orbit, with the number of iterations depending on the specific error configuration. The process is designed to ensure that the RMS orbit deviation at the BPM locations in both planes remains below the analytically computed RMS values~\cite{Dalena_2024} for the given configuration, with a maximum limit of 15 iterations.\\
After this initial optimization, a more precise tuning process is implemented using the response matrix from the HEB model design. This step focuses on correcting key beam parameters, including phase advance, dispersion, coupling, and tunes. 
These quantities are computed at the BPMs using the transfer matrix formalism, assuming no errors in their estimations so far.
To allow corrections based on the response matrix, 560 quadrupole correctors are installed near two (out of six) quadrupole families (the same where the sextupoles are installed), while 568 skew quadrupole correctors are installed close to the sextupoles location. All these correctors operate independently and are not misaligned.\\
The correction process follows a linear relation:
\begin{equation}
    \mathbf{O}^\alpha = \mathbf{A}^\alpha\mathbf{K}
    \label{Eq1}
\end{equation}
where $\mathbf{O}^\alpha$ represents the difference vector of the observable $\alpha$ (such as phase advance,  dispersion, $\ldots$ ) along all BPMs of the ring, comparing the error-affected simulation with the ideal model. $\mathbf{A}^\alpha$ denotes the response matrix associated with the observable $\alpha$, computed across all BPMs and correctors of the lattice under consideration. $\mathbf{K}$ represents the vector of corrector strengths.\\
For computational efficiency, the corrections are divided into two groups: parameters adjusted via the normal quadrupole correctors (such as phase advance in both planes, horizontal-plane dispersion, and both tunes) and parameters primarily corrected by the skew quadrupole correctors (including vertical-plane dispersion and the real and imaginary components of the resonance driving terms $ f_{1001} $ and $ f_{1010} $) \cite{PhysRevAccelBeams.20.054801,PhysRevSTAB.8.034001}.\\
To determine the correctors' strengths needed to reduce the discrepancy between the model and the error-affected simulation, Eq.~\eqref{Eq1} must be inverted. Since the response matrix $\mathbf{A}$ is generally not square, the inversion is performed using the Penrose-Moore pseudo-inverse. Once this is completed, the sextupoles are gradually ramped up, and at each step, four iterations of a loop consisting of one orbit optimization followed by one response matrix optimization are performed.\\
The momentum aperture of the baseline lattice, without errors, reduces to \qty{\pm 0.3}{\percent} when sextupoles are switched off (as shown in Fig.~\ref{fig:momap} for the horizontal plane). At this stage, this value is considered acceptable, since the high energy LINAC can provide dedicated single bunches with an energy spread of \qty{0.05}{\percent}, during commissioning. Further studies will be done to evaluate the dynamic and momentum aperture with errors.

\section{RESULTS}

\begin{figure}[!htb]
   \centering
   \includegraphics[width=.85\columnwidth]{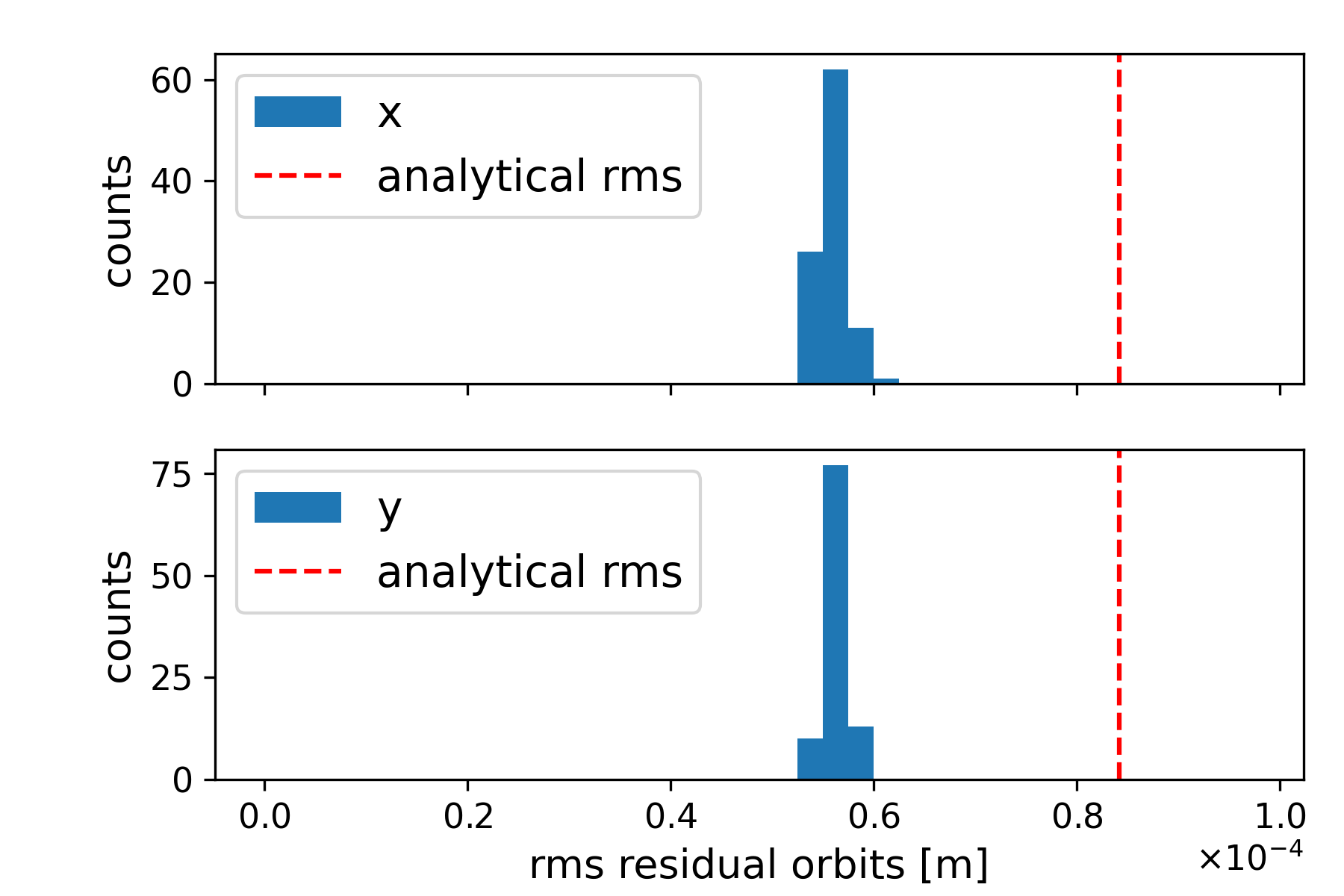}
   \caption{RMS of residual orbit on 100 errors configurations.}
   \label{fig:orbit}
\end{figure}

\begin{figure}[!htb]
   \centering
   \includegraphics[width=\columnwidth]{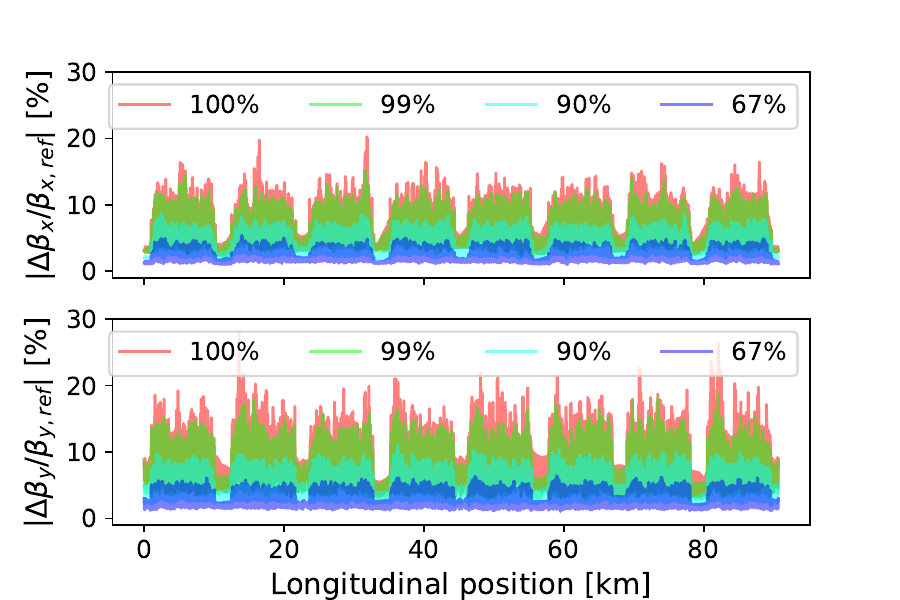}
   \caption{$\beta$-beating quantiles in both planes after the full optimization process.}
   \label{fig:beta}
\end{figure}

\begin{figure}[!htb]
   \centering
    \includegraphics[width=\columnwidth]{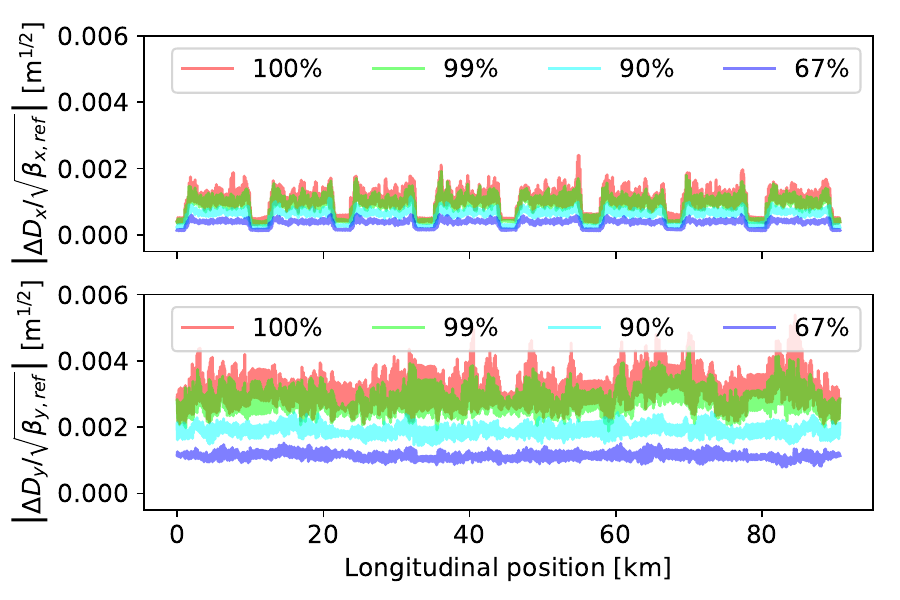}
   \caption{Normalized dispersion quantiles in both planes after the full optimization process.}
   \label{fig:disp}
\end{figure}

\begin{figure}[!htb]
   \centering
   \includegraphics[width=\columnwidth]{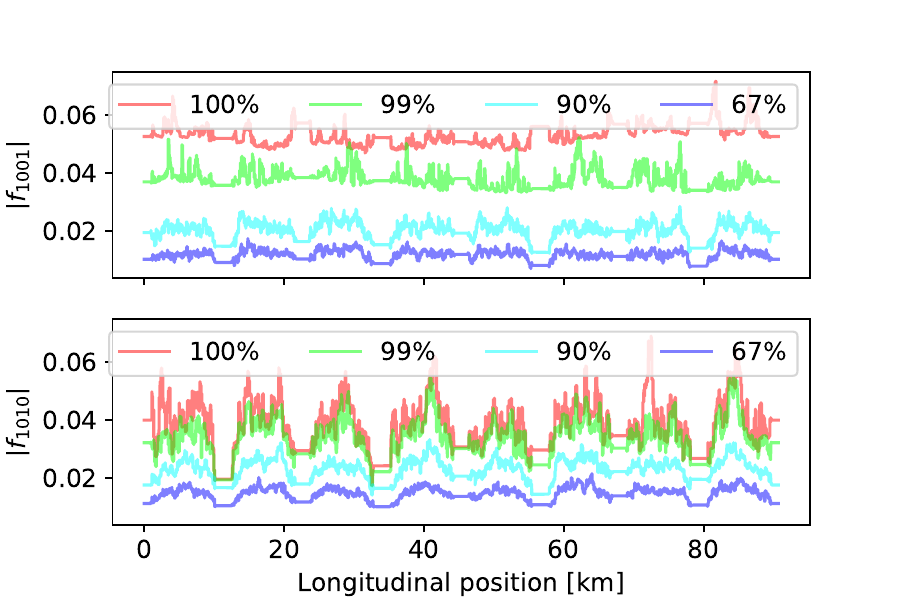}
   \caption{Resonance Driving Terms (RDTs) norms quantiles after the full optimization process.}
   \label{fig:rdts}
\end{figure}

\begin{figure}[!tbh]
    \centering
    \includegraphics[width=\columnwidth]{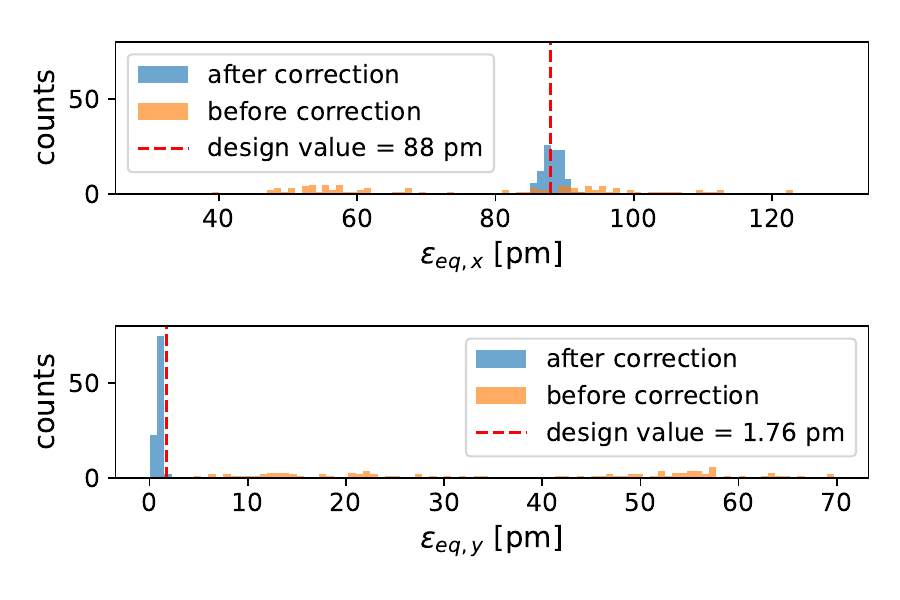}
    \caption{Distribution of the equilibrium emittances in both planes,
    \textit{in orange} before the optimization and
    \textit{in blue} after.}
    \label{fig:emit}
\end{figure}

The correction process described above allows for an RMS residual orbit excursion below the analytical computation, with a distribution quasi-centered around $\approx$~\qty{58}{\um}. The exact distribution across 100 configurations is displayed in Fig.~\ref{fig:orbit}. The corresponding strengths of the orbit correctors remain within the allowed range, with an RMS value below \qty{1}{\milli\tesla\metre} at injection and below \qty{10}{\milli\tesla\metre} during $\rm{t}\rm{\bar{t}}$ operation, as shown in Table~\ref{tab:corr}. 
For the different parameters optimized by the process described previously, the $\beta$-beating is reduced to below \qty{20}{\percent} in both planes (Fig.~\ref{fig:beta}). The normalized dispersion in both planes is also kept under the \qty{2e-3}{\metre\tothe{1/2}} for the $x$ plane, and under \qty{5e-3}{\metre\tothe{1/2}} for the $y$ plane as displayed in Fig.~\ref{fig:disp}. Residual vertical dispersion is more important than residual horizontal dispersion, particularly in the straight sections where no errors are considered. This may indicate the need to add skew quadrupole correctors in the insertions. In addition to these parameters, the coupling is reduced to levels where the resonance driving terms $|f_{1001}|$ and $|f_{1010}|$ are kept well below \qty{10}{\percent} for both planes as seen in Fig.~\ref{fig:rdts}. 
All these optimizations focus on several important parameters, as described in the strategy part of this article, to limit emittance growth in the booster and achieve the target equilibrium emittance in both planes. The target equilibrium emittances are available in \cite{Benedikt:2928793}. For Z operation (\qty{45.6}{\giga\electronvolt}), the targets are \qty{88}{\pm} for the $x$ plane and \qty{1.76}{\pico\metre} for the $y$ plane.
It can be observed that the emittances are well-centered around the design values in both planes, with little spread, as displayed in Fig.~\ref{fig:emit}.
The normal and skew quadrupole correctors used in this process have an RMS integrated value below 1 \unit{\tesla\per\m\m}, as seen in Table~\ref{tab:corr}.
 \begin{table}[!htbp]
	\setlength\tabcolsep{3.5pt}
	\centering
	\caption{Correctors specifications.}
	\label{tab:corr}
	\begin{tabular}{lcc}
		\toprule
		\textbf{Corrector} &  \textbf{E} [\unit{\giga\electronvolt}]  & \textbf{3 $\times$ RMS} \\
		\\
		\midrule
		\textbf{Orbit}     &   20 (inj.)    &   \qty{2.5}{\milli\tesla\m}  \\
		\textbf{X}  &         182.5 ($t\bar{t}$)        &  \qty{23}{\milli\tesla\m}  \\
		\textbf{Orbit}&          20 (inj.)           &    \qty{2.6}{\milli\tesla\m}   \\
		\textbf{Y}  &           182.5 ($t\bar{t}$)        &   \qty{24}{\milli\tesla\m}  \\
		\midrule
		  \textbf{Normal Quad}&  20 (inj.) & \qty{.11}{\tesla\per\m\m} \\
		&            182.5 ($\rm{t}\rm{\bar{t}}$)           &   \qty{.96}{\tesla\per\m\m}   \\
		\textbf{Skew Quad} &                     20 (inj.)   & \qty{.05}{\tesla\per\m\m} \\
		  &                182.5 ($\rm{t}\rm{\bar{t}}$)         &    \qty{.39}{\tesla\per\m\m}  \\
		\bottomrule
	\end{tabular}
\end{table}

%
%

\section{CONCLUSION}

The optimization proposed in this article maintains the equilibrium emittance close to the target values even in the presence of imperfections. This is achieved using approximately \num{4000} correctors, of which about \num{1100} are quadrupole correctors, the rest being orbit correctors. 
Performances can be slightly improved by increasing the number of iterations in the response matrix.
Simulations will be conducted at $\rm{t}\rm{\bar{t}}$ energy (\qty{182.5}{\giga\electronvolt}), incorporating tapering to maintain a stable orbit.
A continuation of this study is undergoing to add other errors contributions such as longitudinal misalignment.

Longer-term improvements can be achieved in order to reduce the number of corrector magnets. In particular, a study of the correlations between different correctors may help limit their number, as there is room available in terms of their strength.
Additionally, optimizing the placement of these correctors to enhance their efficiency could also reduce their overall number.
A non-linear approach to optimize this process can also be explored, for example as done in \cite{huang:ipac2024-thpc32}.

%
%
\ifboolexpr{bool{jacowbiblatex}}%
	{\printbibliography}%
	{%
	%
%
%
%
%
} 
%
%
%
%
\end{document}